\documentclass[aps,prl,a4paper,reprint,10pt,superscriptaddress]{revtex4-2}
\usepackage[utf8]{inputenc}
\usepackage{amsmath}
\usepackage{amssymb}
\usepackage{braket}	
\usepackage{graphicx}
\usepackage{xcolor}
\usepackage{mathrsfs}
\usepackage[colorlinks=true, citecolor=.]{hyperref}
\usepackage{soul}
\usepackage[export]{adjustbox}
\usepackage{physics}

\DeclareMathOperator*{\argmax}{arg\,max}

\definecolor{mypink}{HTML}{C65B80}
\definecolor{myorange}{HTML}{C7733B}
\definecolor{mygreen}{HTML}{77A451}
\definecolor{myblue}{HTML}{6C93C5}
\definecolor{mypurple}{HTML}{9765CA}

\newcommand{\SY}[1]{{\color{black}#1}}

\begin{document}

\title{Generation of large-amplitude squeezed cat states with near-unity efficiency}

\date{\today}

\author{Simon K. Yung}
\email{sksyung@gmail.com}
\affiliation{Department of Quantum Science and Technology, Research School of Physics, The Australian National University, Canberra, ACT 2601, Australia.}

\author{Matthew S. Winnel}
\noaffiliation

\author{Timothy C. Ralph}
\email{ralph@physics.uq.edu.au}
\affiliation{Centre for Quantum Computation and Communication Technology, School of Mathematics and Physics, University of Queensland, St Lucia, Queensland 4072, Australia.}

\author{Jie Zhao}
\email{jie.zhao@anu.edu.au}
\affiliation{Department of Quantum Science and Technology, Research School of Physics, The Australian National University, Canberra, ACT 2601, Australia.}

\begin{abstract}
	The Gottesman--Kitaev--Preskill encoding has emerged as a leading candidate for fault-tolerant quantum computation with continuous variables. For photonic architectures, the major challenge is preparing high-quality resource states, which can be deterministically synthesised from many large-amplitude cat states. Thus far, only modest-sized optical cat states have been prepared experimentally, and the implemented methods are highly probabilistic. We propose an all-optical scheme utilising quantum non-demolition interactions and photon-number measurements to prepare large-amplitude cat states with near-unity probability. Importantly, no particular photon-number outcome is postselected: all outcomes contribute to the accumulated photon number, with the protocol repeated until a required threshold is reached. We demonstrate that the scheme is robust to realistic levels of photon-loss, the dominant source of error in optical systems. Our results highlight the power of active Gaussian operations for state preparation and pave the way for efficient quantum error correction using bosonic codes.
\end{abstract}

\maketitle

 \textit{Introduction}---Quantum computers have the potential to outpace classical computers at certain tasks, with wide-reaching applications~\cite{nielsen_quantum_2010,ladd_quantum_2010}. However, the detrimental effects of noise must be avoided in order to realise this potential. It is thus generally accepted that some form of quantum error correction is necessary~\cite{preskill_beyond_2025}. The physical overhead required to implement quantum error correction poses a hurdle for scaling error-protected computations. Bosonic coding offers a solution to reducing this overhead, by encoding a logical qubit in each bosonic mode. The leading encoding is the Gottesman--Kitaev--Preskill (GKP) code~\cite{gottesman_encoding_2001}, thanks to its performance against loss and Gaussian noise~\cite{albert_performance_2018}, and its relative ease of universal fault-tolerant computation~\cite{baragiola_all-gaussian_2019}. This is especially true for optical systems, where only room temperature operations and measurements are needed.
 
The challenge to implementing the GKP code lies in the preparation of highly non-Gaussian logical basis states (GKP states), which are superpositions of squeezed states on a lattice. This is a notoriously difficult task in photonic platforms. A scalable approach is the GKP breeding protocol, in which several squeezed cat states---superpositions of opposite-phase squeezed coherent states---are interfered~\cite{vasconcelos_all-optical_2010,etesse_proposal_2014,weigand_generating_2018,konno_logical_2024}. This reduces the problem to the preparation of squeezed cat states with sufficiently large amplitude, squeezing, and generation rate. 

Several methods of preparing optical cat states have been studied and demonstrated, relying on the measurement of a subset of modes of a multimode entangled state~\cite{dakna_generating_1997,ourjoumtsev_generating_2006,neergaard-nielsen_generation_2006,ourjoumtsev_generation_2007,wakui_photon_2007,yoshikawa_purification_2017,asavanant_generation_2017,endo_four-photon_2026,takase_generation_2021}. The major limitation of such protocols is their inherent probabilistic nature due to the conditioning measurements, and there is generally a trade-off between the success probability and cat amplitude. This limits the scalability of the GKP breeding protocol, where many large-amplitude cats are required simultaneously. Multiplexing can boost the generation rate, but at the cost of a hardware overhead. 

The low success probabilities of the aforementioned schemes were addressed in Ref.~\cite{winnel_deterministic_2024}, where some of the present authors introduced iterative schemes for preparing cat states that approach determinism as the number of rounds is increased. However, the requirement of large input photon-number states to realise the full potential of the schemes limits their practicality.    

Here, we present a novel scheme for preparing large-amplitude cat states. The basic operating principle is repeated measurement-induced amplification, which probabilistically enlarges the cat state. The repeated scheme, combined with feedforward Gaussian operations, can prepare cat states for fault-tolerant GKP state preparation with probability approaching unity. The scheme can also be adapted to produce large-amplitude squeezed cat states with near-unity probability. We rely on the standard operations that will be required to perform GKP error correction, along with photon-number resolving detectors. 

Our method lowers the hardware requirements for the GKP breeding protocol, providing a more realistic pathway to preparing high-quality optical GKP states. Beyond their use for quantum error correction, our large-amplitude cat states themselves enable quantum error correction using cat codes~\cite{cochrane_macroscopically_1999,schlegel_quantum_2022,hastrup_all-optical_2022}, and also find applications in Heisenberg-limited metrology by making use of the interference fringes in phase-space~\cite{ralph_coherent_2002,munro_weak-force_2002,joo_quantum_2011}. They are also useful for fundamental tests of quantum theory~\cite{sanders_entangled_1992,jeong_quantum_2003,stobinska_violation_2007} and probing decoherence and the quantum-to-classical transition~\cite{deleglise_reconstruction_2008}. 
 	
\begin{figure*}
	\centering
	\includegraphics[width=0.95\linewidth]{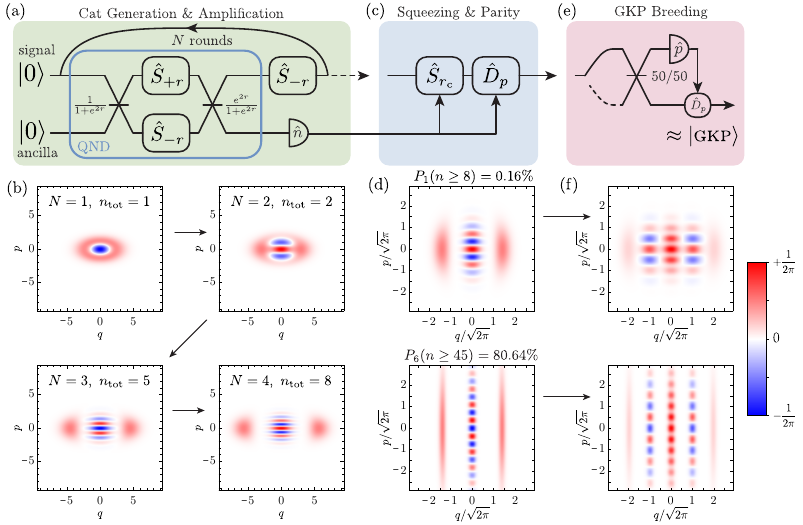}	
	\caption{Preparation of large-amplitude cat states compatible with GKP breeding schemes. (a) Cat generation based on a measurement-induced cat amplifier. In the first round, vacuum signal and ancilla modes are mixed via a QND interaction, after which the signal mode has its $q$-quadrature anti-squeezed. A measurement of $n$ photons in the ancillary mode heralds an approximate cat state $\ket{\mathcal{C}(\sqrt{n},n)}$ at the output. The interaction can be repeated, using a new vacuum ancilla, to amplify the signal cat. After $N$ rounds with $n_\text{tot}$ total detected photons, the final state approximates $\ket{\mathcal{C}(\sqrt{n_\text{tot}},n_\text{tot})}$, independent of the photon-numbers detected in each round.  (b) Wigner functions for an example 4-round amplification series. (c) Conditional squeezing and phase flip. Depending on $n_\text{tot}$, squeezing with $r_c=\ln(n_\text{tot}/\pi)/2$ and displacement $\pi/(4\sqrt{n_\text{tot}})$ in $p$ (if necessary) are applied to the amplified cats in (a), making them useful resources for GKP breeding. (d) Wigner functions for cat states with $\tilde{\alpha}=\sqrt{\pi}$, suitable for one stage of GKP breeding. Upper: given 6 dB squeezers in (a), a squeezed cat with at least 4 dB of squeezing can be prepared with probability $0.16\%$ in a single round ($n_\text{tot}\geq 8$). Lower: the probability can be significantly improved with multiple amplification rounds; after 6 rounds a squeezed cat resource to prepare a GKP state with at least 10 dB symmetric effective squeezing can be prepared with probability $80.64\%$. (e) GKP breeding consumes two squeezed cat states and reduces their $q$-peak spacing by $\sqrt{2}$, but maintains the peak width. Deterministic breeding is enabled by a feedforward displacement. (f) Wigner functions for approximate GKP states produced with two copies of the states in (d). }
	\label{fig:scheme}
\end{figure*}

\textit{Preliminaries}---We target superposition states of the form
\begin{equation}
	\hat{S}(r)\ket{\mathcal{C}(\alpha,t)} \propto \hat{S}(r)\left(\ket{\alpha}+(-1)^t\ket{-\alpha}\right),
\end{equation}
describing a squeezed cat state with real amplitude $\alpha$ and parity $t$. Here, $\ket{\alpha}=\hat{D}(\alpha)\ket{0}$ is a coherent state with displacement $\hat{D}(\alpha) = e^{\alpha\hat{a}^\dagger-\alpha^\ast\hat{a}}$, and $\hat{S}(r)=e^{r(\hat{a}^2-\hat{a}^{\dagger2})/2}$ is the squeezing operator that acts on the field quadratures as $\hat{S}^\dagger(r)\hat{q}\hat{S}(r) = e^{-r}\hat{q}$, $\hat{S}^\dagger(r)\hat{p}\hat{S}(r) = e^{r}\hat{p}$. We use the convention $\hat{q} = \hat{a}+\hat{a}^\dagger$, $\hat{p} = -i(\hat{a}-\hat{a}^\dagger)$, which corresponds to the choice $\hbar=2$~\cite{weedbrook_gaussian_2012}.

In the context of the GKP breeding protocol, the spacing between peaks in the $q$-quadrature needs to be sufficiently large to obtain the correct final grid spacing after several rounds of breeding. In this regard, it is convenient to consider the alternative cat state definition as a superposition of displaced squeezed states: $(\hat{D}(\tilde\alpha)+\hat{D}(-\tilde\alpha))\hat{S}(\tilde r)\ket{0}$, where $4\tilde\alpha$ is the peak spacing in the $q$-quadrature, and $(\tilde\alpha,\tilde r)=(e^{-r}\alpha,r)$. An even cat with $\tilde\alpha=\sqrt{\pi/2}$ serves as a first-approximation to the GKP logical state $\ket{\bar{1}}$, while a cat with $\tilde\alpha =2^{N/2}\sqrt{\pi/2}$ is suitable for $N$ rounds of the GKP breeding protocol, as the peak spacing decreases by $\sqrt{2}$ each round~\footnote{We remark that it is possible to modify the breeding protocol by anti-squeezing each cat by 3 dB before the beamsplitter, so that the peak spacing is preserved. However, in that case, the squeezing of the peaks is reduced instead, and equivalent final states require the same cat amplitude $\alpha$.}. This is our primary motivation for preparing large-amplitude cat states.  

\textit{QND-based cat preparation}---The basic block of our protocol is presented in Fig.~\ref{fig:scheme}(a), which serves as a measurement-induced amplifier for cat states. An input state is combined with an ancillary vacuum mode via a quantum non-demolition (QND) interaction, which performs the transformation~\cite{la_porta_back-action_1989}
\begin{equation}
	\begin{pmatrix}
		\hat{q}_S' \\ \hat{q}_A'
	\end{pmatrix} = \begin{pmatrix}
		1 & 0 \\
		-g & 1
	\end{pmatrix}\begin{pmatrix}
		\hat{q}_S\\ \hat{q}_A
	\end{pmatrix}, \ \begin{pmatrix}
		\hat{p}_S'\\\hat{p}_A'
	\end{pmatrix} = \begin{pmatrix}
		1 & g \\
		0 & 1
	\end{pmatrix}\begin{pmatrix}
		\hat{p}_S \\\hat{p}_A
	\end{pmatrix},
\end{equation}
where $g$ is the QND gain and the subscripts $S$ and $A$ denote the signal and ancilla modes, respectively. The QND interaction can be implemented with beamsplitters and squeezers (see Fig.~\ref{fig:scheme}(a)), in which case the gain is $g=2\sinh{r}$, where $r$ is the squeezing parameter of the inline squeezers. The signal mode is then anti-squeezed and the ancillary mode is measured with a photon-number resolving (PNR) detector. When an initial vacuum signal mode is used, the first round of this scheme was suggested in Ref.~\cite{song_generation_1990} and is similar to the generalized photon subtraction method~\cite{takase_generation_2021} (which does not use the additional squeezer). When $n$ photons are detected, the output closely approximates the state $\ket{\mathcal{C}(\sqrt{n},n)}$.  

We consider performing $N$ amplification rounds by repeatedly feeding the heralded output back to the input. When a cat state $\ket{\mathcal{C}(\sqrt{n},n)}$ is the input of a round, the output heralded on detecting $n'$ photons closely approximates $\ket{\mathcal{C}(\sqrt{n+n'},n+n')}$ such that the input cat is amplified. This amplification can be repeated any number of times and the final cat depends only on the total number of photons detected, $n_\text{tot}$. An example amplification series is shown in Fig.~\ref{fig:scheme}(b). In an ideal system, the fidelity of the output with respect to $\ket{\mathcal{C}(\sqrt{n_\text{tot}},n_\text{tot})}$ scales as $\mathcal{F}\approx 1-0.03/n_\text{tot}$ for $n_\text{tot}\geq 3$ and large squeezing. We refer to this as the \texttt{cat} scheme, and detail the derivation of the output state and $N$-round heralding probability, $P(n_1,\dots,n_N)$, in Supplemental Material I.

With multiple amplification rounds, we can prepare cat states with large, albeit random, amplitude. To make such output states useful, we consider processing them into a squeezed cat state with a particular peak spacing and parity. This can be done by squeezing the output by an amount dependent on the number of detected photons and applying a displacement in $p$ to correct the parity (if necessary), as in Fig.~\ref{fig:scheme}(c). With this in mind, a target peak spacing and minimum peak squeezing can be achieved provided the number of detected photons exceeds a threshold. By performing sufficiently many amplification rounds, we can deterministically prepare resource states for GKP breeding (Fig.~\ref{fig:scheme}(e)). 

We note that by removing the anti-squeezer outside the QND from every round (i.e., the rightmost squeezer in Fig.~\ref{fig:scheme}(a)), a squeezed cat can be prepared directly and the setup is hereafter referred to as the \texttt{squeezed-cat} scheme. This has the practical advantage that it requires fewer inline squeezing operations. The squeezing of the output also increases with the number of rounds, beyond the level of the individual squeezers. However, as we will see shortly, the amplitude of the squeezed cat grows much more slowly than with the third squeezer included, and this simpler scheme is therefore less practical. A similar iterative scheme has also been investigated for directly preparing approximate GKP states~\cite{takase_gottesman-kitaev-preskill_2023}, but offers limited success probability and therefore limited generation rate.   

\textit{Success probability with finite rounds}---With the intention of preparing squeezed cat states with a particular peak spacing, we consider ``success'' to be a detection of at least a threshold of $n_t$ photons. We thus investigate the probability of detecting a total $\SY{n_\text{tot}}\geq n_t$ photons in $N$ rounds, given by 
\begin{equation}
	P_\text{success} = 1-\textstyle\sum_{n_\text{tot} < n_t}P(n_1,\dots,n_N),
\end{equation}
where $n_\text{tot} = \sum_i n_i$. While this sum contains $\binom{n_t+N-1}{N}$ terms, the functional form of each probability $P(n_1,\dots,n_N)$ allows the sum to be recast, using the multinomial expansion, into a manageable form that can be evaluated even for large $n_t$ and $N$ (see Supplemental Material II for details). 

\begin{figure}
	\includegraphics[width=\linewidth]{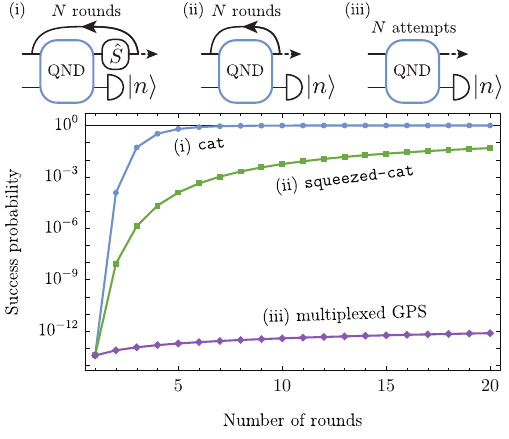}
	\caption{Success probabilities for different schemes: (i) repeated \texttt{cat} amplification; (ii) repeated \texttt{squeezed-cat} amplification; (iii) multiplexed generalised photon subtraction (GPS), where the number of rounds is used to mean the number of instances of GPS, for even comparison. The success threshold is set to $n_t=45$, and 6 dB of squeezing is used.}
	\label{fig:fig2}
\end{figure}

Figure~\ref{fig:fig2} shows a comparison of the success probability of the \texttt{cat} (including squeezer) and \texttt{squeezed cat} (excluding squeezer) schemes for different numbers of rounds. Here, the number threshold is $n_t=45$ such that the output states (see Fig.~\ref{fig:scheme}(d)) can be used for one round of GKP breeding to produce approximate GKP logical states (see Fig.~\ref{fig:scheme}(f)) with at least 10 dB \SY{of symmetric effective} squeezing---a measure recently used to study fault-tolerant thresholds \cite{duivenvoorden_single-mode_2017,aghaee_rad_scaling_2025}. We see that the success probability for the \texttt{cat} scheme rapidly increases with the number of rounds, gaining most of its enhancement within only a few rounds ($N\sim 3$), and significantly outperforms the \texttt{squeezed-cat} scheme. 

For additional comparison, in Fig.~\ref{fig:fig2} we also present the overall success probability for $N$ instances of the generalised photon subtraction (GPS) method, which sees limited improvement via multiplexing. This highlights the source of the improvement in our scheme, where any non-zero photon-number outcomes are useful. Here, we have assumed that 6 dB of squeezing is used throughout the protocols. For larger squeezing, the success probability increases more quickly with respect to the number of rounds, but the general behaviour is preserved (see Supplemental Material).  

\textit{Effect of photon loss---}Practical implementations of the cat amplification schemes will be predominantly affected by photon loss. This can affect the success probabilities and the quality of the output states, which we analyse here. 

We consider pure loss with strength $\ell_\textsc{pnr}$ before the PNR detector, corresponding to an inefficient detection, and $\ell_\textsc{sqz}$ after each squeezer. The squeezing loss can be attributed to finitely squeezed ancillary modes used to implement measurement-induced inline squeezing~\cite{filip_measurement-induced_2005}, or due to mode mismatch \cite{yoshikawa_purification_2017}. In addition to being more realistic, by considering two sources of photon loss, we can distinguish between loss that affects the signal mode and loss that only affects the ancillary mode. 

To determine the effect of loss on the outcome probabilities, we numerically simulate the system using the \texttt{MrMustard} Python package~\footnote{\url{https://github.com/XanaduAI/MrMustard}}. Starting from a vacuum mode, we simulate $N$ amplification rounds, recording the number of detected photons. We then estimate the success probability as the sample success frequency from 5000 repeats. To enable reasonably fast computation, we use a number threshold of $n_t=9$, so that the system is faithfully represented in a 25-dimensional space when 6 dB of squeezing is used. 

\begin{figure}
	\includegraphics[width=\linewidth]{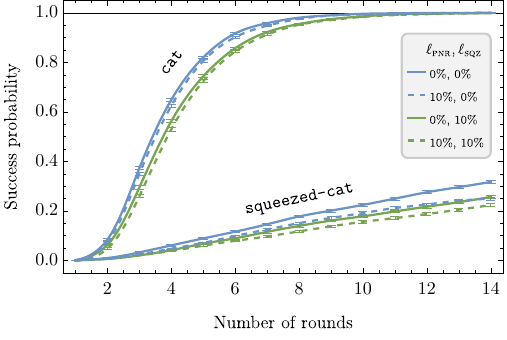}
	\caption{Cat preparation success probabilities including photon loss. Upper: \texttt{cat} protocol; Lower: \texttt{squeezed-cat} protocol. The number threshold is $n_t=9$, and the different line styles denote the probabilities with and without losses associated with the PNR detectors ($\ell_\textsc{pnr}$) and squeezers ($\ell_\textsc{sqz}$). The error bars denote one binomial standard error, $\sqrt{\hat{p}(1-\hat{p})/5000}$, where $\hat{p}$ is the probability estimated from 5000 samples.}
	\label{fig:probability}
\end{figure}

The probabilities including photon loss are shown in Fig.~\ref{fig:probability} for the \texttt{cat} and \texttt{squeezed-cat} schemes. We find that the probabilities are minimally affected, even for relatively large losses (20\% per mode per round). This shows that, even with significant photon loss, large amplitude cats can be grown with high probability even in the finite-round regime. Of course, the success probability will never reach 1, but we can still reach probability $1-\varepsilon$ for small $\varepsilon$ in on the order of 10 rounds in the \texttt{cat} scheme.

When analysing the output states in the presence of photon loss, we find that the fidelity with respect to an ideal cat state depends strongly on the number of detected photons (lower fidelity for higher photon numbers). However, this need not mean that the utility of larger amplitude states is more affected. Therefore, instead of directly studying the fidelity, we investigate the effective amount of loss incurred under different physical losses.
To do so, we maximise the fidelity of the output with respect to an ideal cat state passed through a variable pure loss channel. That is, we determine
\begin{equation}
	\ell_\text{eff} = \argmax_{\ell_\text{eff}, \alpha, t, r_c}\mathcal{F}(\text{output}, L_{\ell_\text{eff}}[\hat{S}(r_c)\ket{\mathcal{C}(\alpha,t)}]),
\end{equation}
where $\mathcal{F} = \left(\Tr\sqrt{\sqrt{\rho_1}\rho_2\sqrt{\rho_1}}\right)^2$ is the fidelity between mixed states $\rho_1$ and $\rho_2$, and $L_{\ell_\text{eff}}[\cdot]$ denotes a pure loss channel of strength $\ell_\text{eff}$. We then study the effective loss across different physical loss parameters, photon-number measurement outcomes, and numbers of rounds. While we can exactly determine the lossy output state for a single round (see Supplemental Material III), an analytical approach is impractical for multiple rounds, so we again numerically simulate the system. 

\begin{figure}
	\centering
	\includegraphics[width=\linewidth]{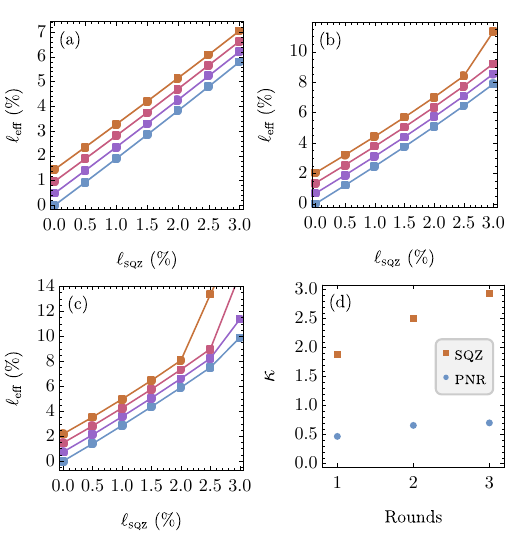}
	\caption{Effective loss of output cat states. (a)--(c) Effective losses after one, two, and three rounds, respectively. The effective loss is presented as a function of the squeezer loss $\ell_\textsc{sqz}$, with trends for different PNR detector loss $\ell_\textsc{pnr}$ in order $\ell_\textsc{pnr}=0.0\%, 1.0\%, 2.0\%, 3.0\%$, from bottom to top. The effective loss is a weighted average over PNR outcomes with up to 9 total photons.     (d) Effective loss coefficients $\kappa_\textsc{pnr}$ and $\kappa_\textsc{sqz}$ for different rounds, fitted from the linear regime up to $\ell_\textsc{sqz}=2\%$.}	\label{fig:effectiveloss}
\end{figure}

We find that with low loss in each round, the effective loss $\ell_\text{eff}$ is approximately constant across photon-number outcomes, and scales linearly with $\ell_\textsc{pnr}$ and $\ell_\textsc{sqz}$: $\ell_\text{eff} \simeq \kappa_\textsc{pnr} \ell_\textsc{pnr}+\kappa_\textsc{sqz} \ell_\textsc{sqz}$. This is demonstrated in Fig.~\ref{fig:effectiveloss}\text{(a)--(c)} for the first three rounds, where the reported effective losses are a weighted average over different outcomes with up to 9 detected photons, where only the most probable $95\%$ and $80\%$ are included for two and three rounds, respectively, to keep the simulations manageable (see Supplemental Material III for details). For multiple rounds, the proportionality factors are increased, though by a factor less than the round number, as depicted in Fig.~\ref{fig:effectiveloss}(d). This demonstrates that it is realistically possible to implement repeated amplification without a significant loss penalty.

The linear trend continues until the effective loss reaches $\sim 8\%$. At this point, there are significant variations in the effective loss across outcomes, leading to an increase in the average effective loss. 

\textit{Experimental feasibility}---Our iterative scheme is compatible with the loop-based architecture for optical quantum computation, wherein an optical pulse is maintained in an optical loop, and operations can be performed using ancillary pulses~\cite{takeda_universal_2017}, augmented with PNR measurements. This is a scalable platform with which several rounds of the scheme can be performed without increasing experimental complexity. Several implementations of the necessary components have been demonstrated~\cite{enomoto_programmable_2021,yonezu_time-domain_2023,okuno_time-domain_2024,yoshida_sequential_2025}. In particular, multiple applications of a squeezing gate have been demonstrated on a cat state, with sufficiently low loss that Wigner negativity was retained after three rounds~\cite{yoshida_sequential_2025}. A universal squeezing gate using noiseless linear amplification has also been shown to squeeze cat states with high fidelity and near-unity success probability~\cite{zhao_high-fidelity_2020}.

As pointed out in Ref.~\cite{takase_gottesman-kitaev-preskill_2023}, a circuit equivalent to Fig.~\ref{fig:scheme}(a) that does not use inline squeezers is possible. The $N$-round circuit is equivalent to preparing a particular $(N+1)$-mode Gaussian state and measuring all but one mode with photon counters. Any such Gaussian state (with zero quadrature means) can be prepared by injecting squeezed vacuum modes into a linear interferometer~\cite{braunstein_squeezing_2005}, which may be simpler to implement. However, the multi-round equivalent circuit requires significantly higher squeezing for one of the input modes (see Supplemental Material IV).
This conversion is similar to existing Gaussian boson sampling approaches to generating non-Gaussian states~\cite{su_conversion_2019,tzitrin_progress_2020}, where a desired state is heralded by a particular PNR outcome pattern. Our scheme has the additional property that different measurement patterns can produce the same state, which boosts success probabilities; the success probability can be further improved by the inclusion of fed-forward inline squeezing.  

We also note that the PNR detection resolution is an important factor. The maximum resolvable photon-number should be greater than any probable detection to avoid heralding a mixed state. For instance, with 6 dB of squeezing and 3 rounds, more than $90\%$ of outcomes have at most 25 photons detected per round, and this percentage drops to $61\%$ for 4 rounds. However, we believe that this will not pose a serious restriction, given that up to 100-photon resolution has been demonstrated~\cite{cheng_100-pixel_2023}. 

\textit{Conclusions}---We have presented an iterative scheme for preparing large-amplitude optical cat states with high probability. The near-deterministic behaviour relies on the fact that any combination of photon-number outcomes can prepare a useful state when combined with feedforward operations after sufficiently many rounds. The difference between the \texttt{cat} and \texttt{squeezed-cat} schemes highlights the power of active squeezing for improving success probabilities, which may find applications in other protocols. We hope that this will encourage improvements in the technical capability of such operations.  

Our proposal is compatible with proven experimental platforms and produces output states that can be used for quantum error correction, and thus represents a promising avenue for achieving fault-tolerant optical quantum computation in the near future.

\textit{Acknowledgements---}We thank Hans Bachor and Jiri Janousek for valuable discussions. S.K.Y. is supported by the Australian Government Research Training Program. This work was partially supported by the Australian Research Council Centre of Excellence for Quantum Computation and Communication Technology (Project No. CE170100012). J.Z. acknowledges support from the Australian Research Council (ARC) Discovery Early Career Research Award (DECRA), Grant No. DE260101046.

\bibliography{bib.bib}

\end{document}

\title{Supplemental Material for\\ \textit{Generation of large-amplitude squeezed cat states with near-unity efficiency}}

\author{Simon K. Yung}
\email{sksyung@gmail.com}
\affiliation{Department of Quantum Science and Technology, Research School of Physics, The Australian National University, Canberra, ACT 2601, Australia.}

\author{Matthew S. Winnel}
\noaffiliation

\author{Timothy C. Ralph}
\email{ralph@physics.uq.edu.au}
\affiliation{Centre for Quantum Computation and Communication Technology, School of Mathematics and Physics, University of Queensland, St Lucia, Queensland 4072, Australia.}

\author{Jie Zhao}
\email{jie.zhao@anu.edu.au}
\affiliation{Department of Quantum Science and Technology, Research School of Physics, The Australian National University, Canberra, ACT 2601, Australia.}

\maketitle

\setcounter{equation}{0}
\setcounter{figure}{0}
\setcounter{table}{0}
\setcounter{page}{1}
\makeatletter
\renewcommand{\theequation}{S\arabic{equation}}
\renewcommand{\thefigure}{S\arabic{figure}}
\renewcommand{\bibnumfmt}[1]{[S#1]}
\renewcommand{\citenumfont}[1]{S#1}

\begingroup
\hypersetup{linkcolor=black}
\tableofcontents
\endgroup

\section*{Conventions and useful properties}

We use natural units with $\hbar=2$, such that the quadrature operators are defined by 
\begin{equation}
	\hat{q} = \hat{a}+\hat{a}^\dagger, \quad \hat{p} = -i(\hat{a}-\hat{a}^\dagger).
\end{equation}
In these units, the vacuum state has quadrature variances $V_q = 1$ and $V_p=1$. In these units, the following wavefunctions, Wigner functions, and properties are useful:
\begin{itemize}
	\item Vacuum state $\ket{0}$: $\phi_0(q) = \frac{1}{(2\pi)^{1/4}}e^{-q^2/4}$ and $W_0(q,p) = \frac{1}{2\pi}e^{-(q^2+p^2)/2}$.
	\item Fock state $\ket{n}$: $\phi_n(q)= \frac{1}{\sqrt{2^n n!}}\left(\frac{1}{2\pi}\right)^{1/4}e^{-\frac{1}{4}q^2}H_n(q/\sqrt{2})$ and $W_n(q,p) = \frac{(-1)^n}{2\pi}e^{-(q^2+p^2)/2}L_n(q^2+p^2)$
	\item The trace of two $N$-mode operators can be expressed using their Wigner functions as
\begin{equation}
	\Tr[\hat{A}\hat{B}] = (4\pi)^N \iint \mathrm{d}^Nq\mathrm{d}^Np~ W_A(q,p)W_B(q,p).
\end{equation}
	\item Similarly, a partial trace over $M$ modes is 
\begin{equation}
	\Tr_M[\hat{A}\hat{B}] = (4\pi)^M \iint \mathrm{d}^Mq\mathrm{d}^Mp~ W_A(q,p)W_B(q,p).
\end{equation}
\end{itemize}

\section{Output states and probabilities excluding losses}
In this section, we detail the calculations of the output states and their associated probabilities for both versions of our scheme. We first detail the calculation for a single round, following Ref.~\cite{takase_generation_2021}. We then extend the calculation to an arbitrary number of rounds. 

\subsection{Single round}
The general process will be to write the wavefunction of the two-mode Gaussian state before the heralding detection, and then calculate the effect of the photon-number measurement. Let $q_0,q_1$ be the $q$-quadratures of the two modes (mode 0 will be the signal mode). A general pure two-mode Gaussian state with zero mean ($\langle q_i\rangle =0$) can be represented by
\begin{equation}
	\ket{G} = \iint\mathrm{d}q_0\mathrm{d}q_1~G(q_0,q_1)\ket{q_0}\ket{q_1},
\end{equation}
with a Gaussian function
\begin{equation}
	G(q_0,q_1) = \frac{1}{(2\pi)^{1/2}(\det V_q)^{1/4}}\exp\left(-\frac{1}{4}\mathbf{q}^\top V_q^{-1}\mathbf{q}\right),
\end{equation}
where $\mathbf{q} = (q_0,q_1)^\top$ and 
\begin{equation}
	V_q = \begin{pmatrix}
		\langle q_0^2\rangle & \langle q_0q_1 \rangle \\
		\langle q_1q_0\rangle & \langle q_1^2\rangle
	\end{pmatrix}
\end{equation}
is the covariance matrix for the $q$-quadratures. It is convenient to express the exponent in terms of the inverse of $V_q$, say $\sigma=V_q^{-1} = \left(\begin{smallmatrix}
	\sigma_{00} & \sigma_{01} \\
	\sigma_{01} & \sigma_{11}
\end{smallmatrix}\right)$. Therefore, we can write
\begin{equation}
	\mathbf{q}^\top \sigma \mathbf{q} = \sigma_{00}q_0^2 +2\sigma_{01}q_0q_1 + \sigma_{11}q_1^2 = \frac{\det\sigma}{\sigma_{11}}q_0^2 + \sigma_{11}\left(\frac{\sigma_{01}}{\sigma_{11}}q_0+q_1\right)^2,
\end{equation}
so that 
\begin{align}
	G(q_0,q_1) &= \frac{(\det\sigma)^{1/4}}{(2\pi)^{1/2}}\exp\left(-\frac{1}{4}\left(\frac{\det\sigma}{\sigma_{11}}q_0^2 + \sigma_{11}\left(\frac{\sigma_{01}}{\sigma_{11}}q_0+q_1\right)^2\right)\right) \\
	&= (\det\sigma)^{1/4} \phi_0\left(\sqrt{\frac{\det\sigma}{\sigma_{11}}}q_0\right)\phi_0\left(\sqrt{\sigma_{11}}\left(\frac{\sigma_{01}}{\sigma_{11}}q_0+q_1\right)\right),
\end{align}
where $\phi_0(q)$ is the $q$-quadrature wavefunction of the vacuum. To model a photon-number detection on mode 1, we project the two-mode state onto the Fock state $\ket{n_1}$ on mode 1:
\begin{equation}
	\bra{n_1}\ket{G} = \iint\mathrm{d}q_0\mathrm{d}q_1~G(q_0,q_1)\braket{n_1}{q_1}\ket{q_0} = \int \mathrm{d}q_0\left[\int \mathrm{d}q_1~G(q_0,q_1)\phi_{n_1}(q_1)\right]\ket{q_0}.
\end{equation}
Therefore, the unnormalised wavefunction of the output (conditioned on $n_1$) is 
\begin{equation}
	\Psi_{n_1}(q_0) = \int \mathrm{d}q_1 G(q_0,q_1)\phi_{n_1}(q_1) = (\det\sigma)^{1/4} \phi_0\left(\sqrt{\frac{\det\sigma}{\sigma_{11}}}q_0\right)\int \mathrm{d}q_1~\phi_0\left(\sqrt{\sigma_{11}}\left(\frac{\sigma_{01}}{\sigma_{11}}q_0+q_1\right)\right)\phi_{n_1}(q_1).
\end{equation}
Note the convolution~\cite{takase_generation_2021}
\begin{align}
	(\phi_0\ast \phi_n)(q) = \int \mathrm{d}y~\phi_0(q-y)\phi_n(y) &= \frac{1}{\sqrt{2\pi}}\frac{1}{\sqrt{2^nn!}}\int \mathrm{d}y~e^{-(q-y)^2/4}e^{-y^2/4}H_n(y/\sqrt{2}) \\
	&= \frac{1}{\sqrt{2\pi}}\frac{1}{\sqrt{2^nn!}} e^{-q^2/8}\int\mathrm{d}y~e^{-(q/2-y)^2/2}H_n(y/\sqrt{2}) \\
	&=  \frac{1}{\sqrt{2^nn!}} e^{-q^2/8}(q/\sqrt{2})^n \\
	&= \frac{1}{2^n\sqrt{n!}}e^{-q^2/8}q^n
\end{align}
Now we specialise to the two schemes:
\subsubsection{\texttt{squeezed-cat} scheme}
Here, we have
\begin{equation}
	V_q = 	\begin{pmatrix}
		-\sqrt{R_2} & \sqrt{1-R_2} \\
		\sqrt{1-R_2} & \sqrt{R_2}
	\end{pmatrix}\begin{pmatrix}
		e^{-2r} & \\
		& e^{2r} 
	\end{pmatrix}\begin{pmatrix}
		-\sqrt{R_2} & \sqrt{1-R_2} \\
		\sqrt{1-R_2} & \sqrt{R_2}
	\end{pmatrix} = \begin{pmatrix}
		1 & 2\sinh(r) \\
		2\sinh(r) & -1+\cosh(2r)
	\end{pmatrix},
\end{equation}
where $R_2 = e^{2r}/(1+e^{2r})$. Then
\begin{equation}
	\sigma = \begin{pmatrix}
		1+4\sinh^2(r) & -2\sinh(r) \\
		-2\sinh(r) & 1 
	\end{pmatrix}, \quad \det\sigma=1,
\end{equation}
so the conditional output wavefunction is
\begin{align}
	\Psi_{n_1}(q_0) &= \phi_0(q_0)\int \mathrm{d}q_1~\phi_0\left(\left(-2\sinh(r)q_0+q_1\right)\right)\phi_{n_1}(q_1) \\
	&= \phi_0(q_0) \frac{1}{2^{n_1}\sqrt{n_1!}}(2\sinh(r)q_0)^{n_1}e^{-\sinh^2(r)q_0^2/2} \\
	&= \frac{1}{(2\pi)^{1/4}\sqrt{n_1!}}(\sinh(r))^{n_1} q_0^{n_1}e^{-\frac{1+2\sinh^2(r)}{4}q_0^2} \label{eq:sqcatwavefunction} \\
	&= \frac{1}{(2\pi)^{1/4}\sqrt{n_1!}}(\sinh(r))^{n_1} q_0^{n_1}e^{-\cosh(2r)q_0^2/4}. 
\end{align}

\subsubsection{\texttt{cat} scheme}
Similarly, for the \texttt{cat} scheme with the additional squeezer, we have
\begin{align}
	V_q &= \begin{pmatrix}
		e^{r} & \\
		& 1
	\end{pmatrix}\begin{pmatrix}
		-\sqrt{R_2} & \sqrt{1-R_2} \\
		\sqrt{1-R_2} & \sqrt{R_2}
	\end{pmatrix}\begin{pmatrix}
		e^{-2r} & \\
		& e^{2r} 
	\end{pmatrix}\begin{pmatrix}
		-\sqrt{R_2} & \sqrt{1-R_2} \\
		\sqrt{1-R_2} & \sqrt{R_2}
	\end{pmatrix}\begin{pmatrix}
		e^{r} & \\
		& 1
	\end{pmatrix} \\
	&= \begin{pmatrix}
		e^{2r} & -1+e^{2r} \\
		-1+e^{2r} & -1+e^{-2r}+e^{2r}
	\end{pmatrix}.
\end{align}
Then 
\begin{equation}
	\sigma = \begin{pmatrix}
		1+e^{-4r}-e^{-2r} & -1+e^{-2r} \\
		-1+e^{-2r} & 1
	\end{pmatrix}, \quad \det\sigma = e^{-2r},  
\end{equation}
so the conditional output wavefunction is
\begin{align}
	\Psi_{n_1}(q_0) &= (e^{-2r})^{1/4} \phi_0\left(e^{-r}q_0\right)\int \mathrm{d}q_1~\phi_0\left(\left((-1+e^{-2r})q_0+q_1\right)\right)\phi_{n_1}(q_1) \\
	&= e^{-r/2} \phi_0\left(e^{-r}q_0\right)q_0^{n_1} \frac{1}{2^{n_1}\sqrt{n_1!}}(1-e^{-2r})^{n_1}e^{-(1-e^{-2r})^2q_0^2/8} \\
	&= \frac{e^{-r/2}}{(2\pi)^{1/4}\sqrt{n_1!}}(\sinh(r))^{n_1}(e^{-r}q_0)^{n_1}e^{-(1+e^{-4r})q_0^2/8} \\
	&= \frac{e^{-r/2}}{(2\pi)^{1/4}\sqrt{n_1!}}(\sinh(r))^{n_1}(e^{-r}q_0)^{n_1}e^{-\cosh(2r)e^{-2r}q_0^2/4}.
\end{align}

The similarity of these output states to ideal cat states will be discussed in Sec.~\ref{sec:fidelity}. 

\subsubsection{Single round probabilities}
The heralding probabilities for the first round are identical for both schemes (the additional squeezing is independent of the photon-number detection). The probability of detecting $n$ photons is 
\begin{align}
	P(n) = \int \mathrm{d}q_0~|\Psi_n(q_0)|^2 &= \int\mathrm{d}q_0~\frac{1}{\sqrt{2\pi}n!}(\sinh(r))^{2n} q_0^{2n}e^{-\cosh(2r)q_0^2/2} \\
	&= \frac{2^n}{\sqrt{\pi}n!}\Gamma\left(n+\frac{1}{2}\right)(\sinh(r))^{2n}(\cosh(2r))^{-n-\frac{1}{2}}.
\end{align}

\subsection{Extension to an arbitrary number of rounds}

We begin with an example of two rounds with the \texttt{squeezed-cat} scheme and will proceed to generalise this. We model the system as a series of QND interactions between the signal and two ancillary modes, followed by photon-number measurements on the ancillary modes. The covariance matrix of the system before photon-number measurement is $\sigma^{-1} = OO^\top$ with 
\begin{equation}
	O = \begin{pmatrix}
		-\sqrt{R'} & 0 & \sqrt{1-R'} \\
		0 & 1 & 0 \\
		\sqrt{1-R'} & 0 & \sqrt{R'}
	\end{pmatrix}\begin{pmatrix}
		e^{-r} & 0 & 0 \\
		0 & 1 & 0 \\
		0 & 0 & e^{r}
	\end{pmatrix}\begin{pmatrix}
		\sqrt{R} & 0 & \sqrt{1-R} \\
		0 & 1 & 0 \\
		-\sqrt{1-R} & 0 & \sqrt{R}
	\end{pmatrix}\begin{pmatrix}
		-\sqrt{R'} & \sqrt{1-R'} & 0 \\
		\sqrt{1-R'} & \sqrt{R'} & 0 \\
		0 & 0 & 1
	\end{pmatrix}
	\begin{pmatrix}
		e^{-r} & 0 & 0 \\
		0 & e^r & 0 \\
		0 & 0 & 1
	\end{pmatrix}.
\end{equation}

Then, with $R = \frac{1}{1+e^{2r}}$ and $R' = \frac{e^{2r}}{1+e^{2r}}$, we get
\begin{equation}
	\sigma = \begin{pmatrix}
		1+8\sinh^2{(r)} & 2\sinh{(r)} & -2\sinh{(r)} \\
		2\sinh{(r)} & 1 & 0 \\
		-2\sinh{(r)} & 0 & 1 
	\end{pmatrix}, 
\end{equation}
which we can use to write the three-mode Gaussian function as
\begin{equation}
	G(q_0,q_1,q_2) = \phi_0(q_0) \phi_0(q_1 +2\sinh{(r)}q_0)\phi_0(q_2 - 2\sinh{(r)} q_0).
\end{equation}
Conveniently, the $q_1$ and $q_2$ terms are decoupled, and so the integration for the conditional output is straightforward:
\begin{align}
	\Psi_\mathbf{n}(q_0) &= \phi_0(q_0) \left(\int \mathrm{d}q_1 \ \phi_0(-2\sinh{(r)}q_0 -q_1) \phi_{n_1}(q_1)\right)\left(\int \mathrm{d}q_2 \ \phi_0(2\sinh{(r)}q_0 -q_2) \phi_{n_2}(q_2)\right) \\
	&= \frac{1}{\pi^{1/4}} \frac{1}{\sqrt{2^{n_1} n_1!}\sqrt{2^{n_2}n_2!}} (-2\sinh{(r)})^{n_1}(2\sinh{(r)})^{n_2}q_0^{n_1+n_2}e^{-\frac{1}{4}(2+8\sinh^2{(r)})q_0^2}.
\end{align}
This wavefunction has a similar form to Eq.~\eqref{eq:sqcatwavefunction}; that is, the output approximates a squeezed cat \\ 
$\hat{S} (r_c) \ket{\text{cat}_{\sqrt{n_1+n_2},n_1+n_2}}$, with $e^{2r_c} = 2+8\sinh^2{(r)}$.

Similarly, $N$ rounds of the scheme can be modelled as a series of QND interactions between the signal mode and $N$ ancillary modes, followed by photon-number measurements on the ancillary modes. We can therefore extend the above calculations to an $(N+1)$-mode Gaussian state. The relevant generalisations are
\begin{equation}
	G(q_0,q_1,\dots,q_N) = G(\mathbf{q}) = \frac{(\det\sigma)^{1/4}}{(2\pi)^{(N+1)/4}}\exp\left(-\frac{1}{4}\mathbf{q}^\top \sigma \mathbf{q}\right), 
\end{equation}
and
\begin{equation}
	\Psi_\mathbf{n}(q_0) = \int \mathrm{d}q_1\cdots\mathrm{d}q_N~G(\mathbf{q})\phi_{n_1}(q_1)\cdots\phi_{n_N}(q_N),
\end{equation}
where $\mathbf{n} = (n_1,\dots,n_N)$ denotes the number of photons detected in round (i.e., in each of the ancillary modes). This integral is reasonably simple to evaluate because the form of $\sigma$ for our schemes is such that $G(\mathbf{q})$ can be decomposed into a product of vacuum wavefunctions, each coupling the signal mode to an ancillary mode. The result is a product of single-round wavefunctions. 

\subsubsection{\texttt{squeezed-cat} scheme output state and probabilities}
To calculate the matrix $\sigma$, it is useful to define the following $(N+1)\times(N+1)$ matrices 
\begin{equation}
	M_i = \begin{pmatrix}
		-1 & & & -m & &  \\
		 & 1 & & & & \\
		 & & \ddots & & & \\
		 & & & 1 & & \\
		 & & & & \ddots & \\
		 & & & & & & 1
	\end{pmatrix}, \quad m=2\sinh{(r)},
\end{equation}
where $-m$ is in the $(i+1)$\textsuperscript{th} column. This matrix is precisely the transformation of $\sigma$ from the $(i-1)$\textsuperscript{th} round to the $i$\textsuperscript{th} round: $\sigma_i = M_i\sigma_{i-1}M_i^\top$. The matrix $\sigma$ can then be built, starting from the identity, noting that $\sigma_0$ is the identity because we start with vacuum. The result is
\begin{equation}
	\sigma_N = M_N\cdots M_1 M_1^\top \cdots M_N^\top = \begin{pmatrix}
		1+Nm^2 & (-1)^Nm & (-1)^{N-1}m & \cdots & -m \\
		(-1)^Nm & 1 & 0 & \cdots & 0 \\
		(-1)^{N-1}m & 0 & 1 & \cdots & 0 \\
		\vdots & \vdots & \vdots & \ddots & \vdots \\
		-m & 0 & 0 & \cdots  & 1
	\end{pmatrix}
\end{equation}
The determinant of $\sigma$ is always $|\sigma| = 1$, and we can write $\mathbf{q}^\top \sigma \mathbf{q}$ as 
\begin{equation}
	\mathbf{q}^\top \sigma \mathbf{q} = q_0^2 + \left(q_1+(-1)^N2\sinh{(r)}q_0\right)^2 + \left(q_2+(-1)^{N-1}2\sinh{(r)}q_0\right)^2 + \cdots + \left(q_N+(-1)^{N-(N-1)}2\sinh{(r)}q_0\right)^2. 
\end{equation} 
The $(N+1)$-mode Gaussian function can then be expressed as
\begin{align}
	G(\mathbf{q}) &= \phi_0\left(q_0\right)\phi_0\left(q_1+(-1)^N2\sinh{(r)}q_0\right)\phi_0\left(q_2+(-1)^{N-1}2\sinh{(r)}q_0\right)\cdots \phi_0\left(q_N+(-1)^{N-(N-1)}2\sinh{(r)}q_0\right) \\
	&= \phi_0(q_0) \prod_{i=1}^N \phi_0\left(q_i + (-1)^{N+1-i}2\sinh{(r)}q_0\right).
\end{align}
Finally, the output state unnormalised wavefunction can be expressed as 
\begin{align}
	\Psi_{\mathbf{n}}(q_0) &= \frac{1}{(2\pi)^{1/4}}e^{-q^2/4} \prod_{i=1}^N \frac{1}{2^{n_i}\sqrt{n_i!}}\left((-1)^{N+1-i}\right)^{n_i}\left(2\sinh{(r)}\right)^{n_i}q_0^{n_i}e^{-\frac{1}{8}4\sinh^2{(r)}q_0^2} \\
	&= \left(\frac{1}{(2\pi)^{1/4}}\prod_{i=1}^N \frac{1}{\sqrt{n_i!}}\left((-1)^{N+1-i}\right)^{n_i} \right) \left(\sinh{(r)}\right)^{n_\text{tot}} q_0^{n_\text{tot}} e^{-\frac{1+2N\sinh^2{(r)}}{4}q_0^2},
\end{align}
where $n_\text{tot} = \sum_{i=1}^N n_i$, the total number of photons detected. 

To calculate the output probabilities, we use the integral
\begin{equation}
	\int \mathrm{d}q_0~q_0^{2n} e^{-\frac{1+2N\sinh^2(r)}{2}q_0^2} = 2^{n+\frac{1}{2}}\Gamma\left(n+\frac{1}{2}\right)\left(1+2N\sinh^2(r)\right)^{-n-\frac{1}{2}},
\end{equation}
so that 
\begin{align}
	P_\text{sqzcat}(\mathbf{n}) &= \frac{1}{\sqrt{\pi}}\left(\prod_{i=1}^N\frac{1}{n_i!}\right) (\sinh(r))^{2n_\text{tot}}2^{n_\text{tot}}\Gamma\left(n_\text{tot}+\frac{1}{2}\right)\left(1+2N\sinh^2(r)\right)^{-n_\text{tot}-\frac{1}{2}}.
\end{align}
The probability of not generating a squeezed cat is
\begin{equation}
	P_\text{sqzcat}(\mathbf{0}) = \frac{1}{\sqrt{1+2N\sinh^2(r)}}.
\end{equation}

\subsubsection{\texttt{cat} scheme output state and probabilities}

With the additional squeezer on the signal mode, we have
\begin{equation}
	M_i = \begin{pmatrix}
		-e^{-r} & & & -e^{-r}m & &  \\
		 & 1 & & & & \\
		 & & \ddots & & & \\
		 & & & 1 & & \\
		 & & & & \ddots & \\
		 & & & & & & 1
	\end{pmatrix}, \quad m=2\sinh{(r)},
\end{equation}
and
\begin{equation}
	\sigma_N = M_N\cdots M_1 M_1^\top \cdots M_N^\top = \begin{pmatrix}
		1+e^{-2(N+1)r}-e^{-2r} & (-e^{-r})^Nm & (-e^{-r})^{N-1}m & \cdots & -m \\
		(-e^{-r})^Nm & 1 & 0 & \cdots & 0 \\
		(-e^{-r})^{N-1}m & 0 & 1 & \cdots & 0 \\
		\vdots & \vdots & \vdots & \ddots & \vdots \\
		-m & 0 & 0 & \cdots  & 1
	\end{pmatrix}.
\end{equation}
Now $\det\sigma=e^{-2Nr}$, and we can write
\begin{equation}
	\mathbf{q}^\top \sigma \mathbf{q} = e^{-2Nr}q_0^2 + \left(q_1+(-e^{-r})^N2\sinh{(r)}q_0\right)^2 + 
	\cdots + \left(q_N+(-e^{-r})^{N-(N-1)}2\sinh{(r)}q_0\right)^2.
\end{equation}
Then
\begin{equation}
	G(\mathbf{q}) = e^{-Nr/2}\phi_0(e^{-Nr}q_0)\prod_{i=1}^N \phi_0(q_i+(-e^{-r})^{N+1-i}2\sinh{(r)}q_0),
\end{equation}	
and the unnormalised output wavefunction is
\begin{align}
	\Psi_\mathbf{n}(q_0) &= \frac{e^{-Nr/2}}{(2\pi)^{1/4}}e^{-e^{-2Nr}q_0^2/4}\prod_{i=1}^N \frac{1}{2^{n_i}\sqrt{n_i!}}\left((-e^{-r})^{N+1-i}\right)^{n_i}(2\sinh(r))^{n_i}q_0^{n_i}e^{-\frac{4e^{-2(N+1-i)r}\sinh^2(r)}{8}q_0^2} \\
	&= \frac{e^{-Nr/2}}{(2\pi)^{1/4}}\left(\prod_{i=1}^N \frac{1}{\sqrt{n_i!}}\left((-e^{-r})^{N+1-i}\right)^{n_i}\right) (\sinh(r))^{n_\text{tot}}q_0^{n_\text{tot}} e^{-\frac{1-e^{-2r}+e^{-2Nr}+e^{-2(N+1)r}}{8}q_0^2}. \label{eq:catwavefunction}
\end{align}
To calculate the output probabilities, we now use the integral
\begin{equation}
	\int\mathrm{d}q_0~q_0^{2n} e^{-\frac{1-e^{-2r}+e^{-2Nr}+e^{-2(N+1)r}}{4}q_0^2} = 2^{2n+1}\Gamma\left(n+\frac{1}{2}\right) \left(1-e^{-2r}+e^{-2Nr}+e^{-2(N+1)r}\right)^{-n-\frac{1}{2}},
\end{equation}
which gives
\begin{align}
	P_\text{cat}(\mathbf{n}) = \frac{e^{-Nr}}{\sqrt{\pi}}\left(\prod_{i=1}^N \frac{\left((e^{-2r})^{N+1-i}\right)^{n_i}}{n_i!}\right)(\sinh(r))^{2n_\text{tot}}2^{2n_\text{tot}+\frac{1}{2}}\Gamma\left(n_\text{tot}+\frac{1}{2}\right) \left(1-e^{-2r}+e^{-2Nr}+e^{-2(N+1)r}\right)^{-n_\text{tot}-\frac{1}{2}}. \label{eq:catprob}
\end{align}
For this scheme, the probability of not generating a cat is
\begin{equation}
	P_\text{cat}(\mathbf{0}) = \frac{\sqrt{2}e^{-Nr}}{\sqrt{1-e^{-2r}+e^{-2Nr}+e^{-2(N+1)r}}}.
\end{equation}

\subsection{Output state fidelity with respect to ideal cat} \label{sec:fidelity}

The normalised wavefunction of the output after detecting $n$ photons is 
\begin{equation}
	\Psi_n(q) = \frac{q^n \exp\left(-\cosh(2r)\exp(-2r)q^2/4\right)}{2^n \sqrt{2(1+\exp(-4r))^{-n-1/2}\Gamma(n+1/2)}}\approx \frac{q^n \exp\left(-q^2/8\right)}{2^n \sqrt{2\Gamma(n+1/2)}},
\end{equation}
where the approximation is valid for large squeezing (large $r$). On the other hand, the wavefunction of a cat $\ket{\mathcal{C}(\sqrt{n},n)}$ is 
\begin{equation}
	\Psi_{\mathcal{C}(\sqrt{n},n)}(q) = \frac{\exp\left(-\frac{(q+2\sqrt{n})^2}{4}\right)+(-1)^n \exp\left(-\frac{(q-2\sqrt{n})^2}{4}\right)}{2^{3/4}\pi^{1/4}\sqrt{1+(-1)^n e^{-2n}}}.
\end{equation}
We can calculate the fidelity via
\begin{equation}
	\mathcal{F} = \braket{\Psi_n}{\mathcal{C}(\sqrt{n},n)}^2 = \left(\int \mathrm{d}q~\Psi_n(q)\Psi_{\mathcal{C}(\sqrt{n},n)}(q)\right)^2,
\end{equation}
which gives 

\begin{equation}
    \mathcal{F} = \frac{2^{n+1/2}e^{-4(n-2)r}\left(1+e^{-4r}\right)^{n+1/2}\left(3+e^{-4r}\right)^{-n}}{\sqrt{\pi } \left((-1)^n+e^{2 n}\right) \left(3 e^{4 r}+1\right)^2 \Gamma \left(n+\frac{1}{2}\right)} A \label{eq:fidelity}
\end{equation}
with 
\begin{multline}
    A = \left((-1)^n+1\right)^2 \left(3 e^{4 r}+1\right) e^{4 (n-1) r} \Gamma \left(\frac{n+1}{2}\right)^2 \, _1F_1\left(\frac{n+1}{2};\frac{1}{2};\frac{2 e^{4 r} n}{1+3 e^{4 r}}\right){}^2 \\+8 \left((-1)^n-1\right)^2 n e^{4 n r} \Gamma \left(\frac{n}{2}+1\right)^2 \, _1F_1\left(\frac{n}{2}+1;\frac{3}{2};\frac{2 e^{4 r} n}{1+3 e^{4 r}}\right){}^2
\end{multline}
where $_1F_1(a;b;z)$ is a confluent hypergeometric function. 

\begin{figure}
	\centering
	\includegraphics{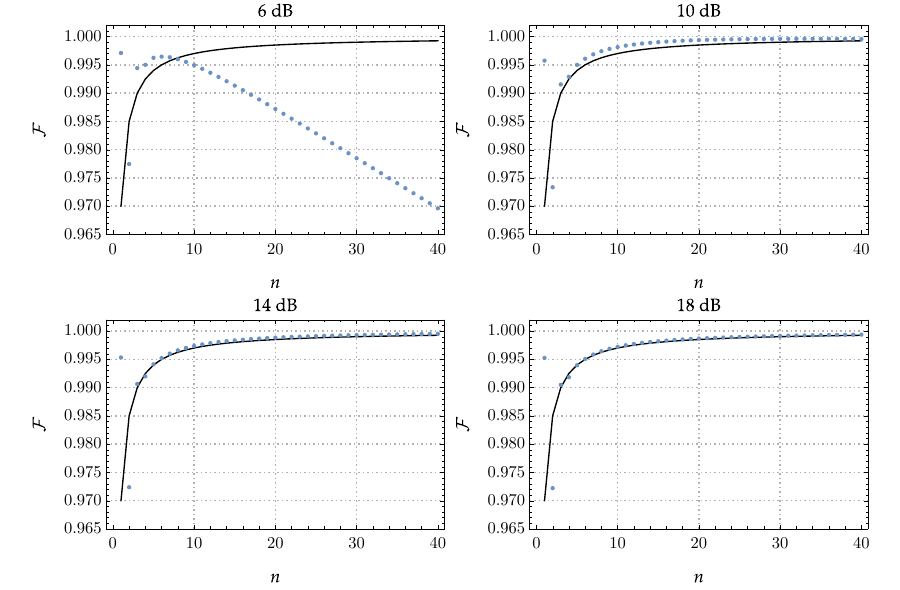}	
	\caption{Fidelity of output states with respect to a cat state $\ket{\mathcal{C}(\sqrt{n},n)}$, for different squeezing levels. The black line is $F=1-0.03/n$, which approximates the fidelity for $n>0$ and large squeezing. The blue points are calculated from Eq.~\eqref{eq:fidelity}}
	\label{fig:fidelityn}
\end{figure}

The fidelity for different values of $n$ is plotted in Fig.~\ref{fig:fidelityn}, for different squeezing levels. For large squeezing, the fidelity scales as roughly $1-0.03/n$, as found in Ref.~\cite{takase_generation_2021}. 

\section{Thresholded success probabilities}
From Eq.~\eqref{eq:catprob}, we have that the probability of detection numbers $\mathbf{n}=(n_1,\dots,n_N)$ has the form
\begin{equation}
	P_\text{cat}(\mathbf{n}) = \underbrace{\frac{e^{-Nr}}{\sqrt{\pi}} (\sinh(r))^{2n_\text{tot}}2^{2n_\text{tot}+\frac{1}{2}}\Gamma\left(n_\text{tot}+\frac{1}{2}\right) \left(1-e^{-2r}+e^{-2Nr}+e^{-2(N+1)r}\right)^{-n_\text{tot}-\frac{1}{2}}}_{f(N,n_{tot})}\prod_{i=1}^N\frac{a_i^{n_i}}{n_i!},
	\end{equation}
	with $a_i=(e^{-2r})^{N+1-i}$.

Setting a success threshold number of photons $n_t$, we are interested in the value of 
\begin{equation}
	P_\text{success}(n_t) = \sum_{n_1+\cdots+n_N\geq n_t} P_\text{cat}(\mathbf{n}) = 1-\sum_{n_1+\cdots+n_N< n_t} P_\text{cat}(\mathbf{n}).
\end{equation}
Na\"ively, to evaluate this probability, we must sum over all possible combinations of detection numbers that sum to less than the threshold. For even moderate numbers of rounds and moderate values of $n_t$, this becomes a prohibitively large calculation. However, the summation can be rearranged as
\begin{equation}
	P_\text{success}(n_t) =1-\sum_{s=0}^{n_t-1}\sum_{n_1+\cdots+n_N=s} P_\text{cat}(\mathbf{n}) = 1-\sum_{s=0}^{n_t-1}f(N,s)\sum_{n_1+\cdots+n_N=s} \prod_{i=1}^N\frac{a_i^{n_i}}{n_i!}.
\end{equation}
Now, the multinomial theorem gives that (for $k_i\geq 0$)
\begin{equation}
	\sum_{n_1+n_2+\cdots+n_N=s}\frac{s!}{n_1!n_2!\cdots n_N!}a_1^{n_1}a_2^{n_2}\cdots a_N^{n_N} = (a_1+a_2+\cdots +a_N)^s,
\end{equation}
which removes the need to explicitly sum over all combinations. That is, we can make the simplification
\begin{equation}
	P_\text{success}(n_t) =1-\sum_{s=0}^{n_t-1}\frac{f(N,s)}{s!}\left(\sum_{i=1}^N (e^{-2r})^{N+1-i}\right)^s,
\end{equation}
which can be simply evaluated. A similar calculation can be achieved for the \texttt{squeezed-cat} scheme. 

\begin{figure}
	\centering
	\includegraphics{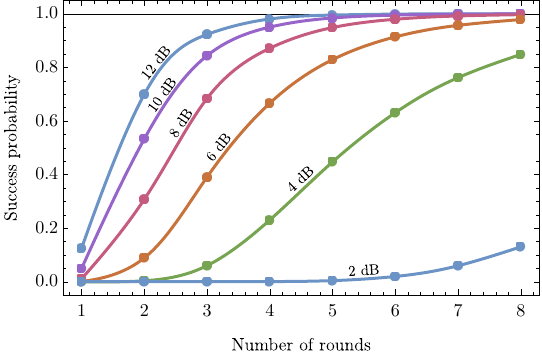}
	\caption{Thresholded cat preparation success probabilities (lossless) with different levels of squeezing. The success probabilities are the probabilities of preparing a cat with amplitude at least $\alpha=3$, determined by detecting at least 9 photons.}
	\label{fig:prob-squeeze}
\end{figure}

\begin{figure}
	\includegraphics{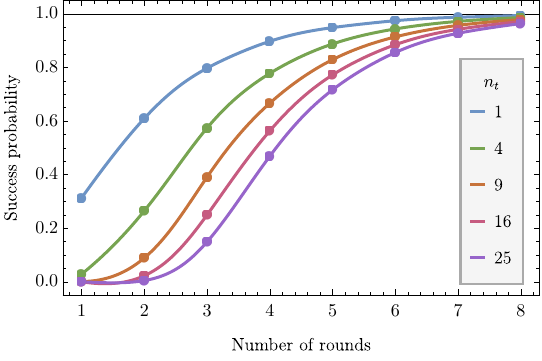}
	\caption{Thresholded cat preparation success probabilities (lossless) with different thresholds. The success probabilities are the probabilities of preparing a cat with amplitude at least $\alpha=\sqrt{n_t}$, determined by detecting at least $n_t$ photons, and calculated based on the exact detection probability with 6 dB squeezing.}
	\label{fig:prob-threshold}
\end{figure}

In Fig.~\ref{fig:prob-squeeze}, we show the effect of the squeezing level on the thresholded success probabilities. This means that a certain success probability can be achieved in fewer rounds by increasing the squeezing level. Similarly, in Fig.~\ref{fig:prob-threshold}, we show how the success probabilities depend on the threshold number of photons.

\subsection{Choosing a suitable number threshold}

We consider preparing approximate GKP logical states with the breeding protocol, using the output from our \texttt{cat} scheme as inputs. To quantify the quality of the output states, we use the effective squeezing~\cite{duivenvoorden_single-mode_2017} of GKP logical qubits defined as
\begin{equation}
    \Delta_q^2 = \frac{-2}{|\beta_q|^2}\ln \left(|S_q|\right) = -\frac{\ln(|S_q|)}{\pi}, \quad \Delta_p^2 = \frac{-2}{|\beta_p|^2}\ln\left(|S_p|\right) = -\frac{\ln(|S_p|)}{\pi},
\end{equation}
where $S_q = \langle D(\beta_q)\rangle$ and $S_p = \langle D(\beta_p)\rangle$, with $\beta_q = i\sqrt{2\pi}$ and $\beta_p=\sqrt{2\pi}$. These values, together with the symmetric squeezing $\Delta_\text{sym}^2 = (\Delta_q^2+\Delta_p^2)/2$, quantify how close the state is to a simultaneous eigenvalue of the stabilizers $D(\beta_q)$ and $D(\beta_p)$~\cite{aghaee_rad_scaling_2025}.  

$D(\alpha) = e^{\alpha a^\dagger -\alpha^* a}$, and $q=a+a^\dagger$, $p = -i(a-a^\dagger)$, so $a = (q+ip)/2, a^\dagger =(q-ip)/2$, and $D(\alpha) = e^{\alpha(q-ip)/2-\alpha^*(q+ip)/2} = e^{(\alpha-\alpha^*) q/2 -i(\alpha+\alpha^*)p/2} = e^{i\Im(\alpha)q-i\Re(\alpha)p}$. Then $D(\sqrt{2\pi}) = e^{-i\sqrt{2\pi}p}$ and $D(i\sqrt{2\pi}) = e^{i\sqrt{2\pi}q}$. 

For GKP breeding relying on homodyne measurements of the $p$-quadrature, it is convenient to work with the $p$-wavefunction. To see this, consider two copies of an input state $\ket{\tilde\psi} = \int\mathrm{d}p~\tilde{\psi}(p)\ket{p}$ combined on a 50/50 beamsplitter followed by measuring the $p$-quadrature of the second mode:
\begin{align}
    \ket{\tilde{\psi}}_1\ket{\tilde{\psi}}_2  &= \int \mathrm{d}p_1\mathrm{d}p_2~\tilde{\psi}(p_1)\tilde{\psi}(p_2)\ket{p_1}\ket{p_2} \\
    &\underset{\text{BS}}{\rightarrow} \int\mathrm{d}p_1\mathrm{d}p_2~\tilde{\psi}\left(\frac{p_1+p_2}{\sqrt{2}}\right)\tilde{\psi}\left(\frac{p_1-p_2}{\sqrt{2}}\right)\ket{p_1}\ket{p_2}\\
    &\underset{\bra{p_2=p_\text{meas}}}{\rightarrow}  \int\mathrm{d}p_1\mathrm{d}p_2~\tilde{\psi}\left(\frac{p_1+p_2}{\sqrt{2}}\right)\tilde{\psi}\left(\frac{p_1-p_2}{\sqrt{2}}\right)\delta\left(p_2-p_\text{meas}\right)\ket{p_1} \\
    &= \int\mathrm{d}p_1~\tilde{\psi}\left(\frac{p_1+p_\text{meas}}{\sqrt{2}}\right)\tilde{\psi}\left(\frac{p_1-p_\text{meas}}{\sqrt{2}}\right)\ket{p_1}.
\end{align}
That is, the post-selected (unnormalised) wavefunction is $\tilde{\psi}_{p_\text{meas}}(p) = \tilde{\psi}\left(\frac{p+p_\text{meas}}{\sqrt{2}}\right)\tilde{\psi}\left(\frac{p-p_\text{meas}}{\sqrt{2}}\right)$. For the following analysis, we assume $p_\text{meas}=0$. Using this formalism, from the $p$-wavefunction of the approximate cat states, we can simply calculate the wavefunction of output approximate GKP states. The expectation values of the stabilizers can also be calculated directly from the $p$-wavefunction of the output, $\tilde{\psi}_\text{GKP}$:
\begin{align}
S_p = \langle D(\beta_p)\rangle &= 
    \bra{\tilde{\psi}_\text{GKP}}e^{-i\sqrt{2\pi}p}\ket{\tilde{\psi}_\text{GKP}} \\
    &= \int \mathrm{d}p~\tilde{\psi}_\text{GKP}^*(p)\tilde{\psi}_\text{GKP}(p)e^{-i\sqrt{2\pi}p},\\
    S_q = \langle D(\beta_q)\rangle &= \bra{\tilde{\psi}_\text{GKP}}e^{i\sqrt{2\pi}q}\ket{\tilde{\psi}_\text{GKP}} \\
    &= \int \mathrm{d}p~\tilde{\psi}_\text{GKP}^*(p)\tilde{\psi}_\text{GKP}(p-2\sqrt{2\pi}). 
\end{align}

The approximate cat states prepared via the $N$-round \texttt{cat} scheme have unnormalised wavefunction (see Eq.~\eqref{eq:catwavefunction})
\begin{equation}
    \Psi_\textbf{n}(q) = q^{n_\text{tot}}e^{-\frac{1-e^{-2r}+e^{-2Nr}+e^{-2(N+1)r}}{8}q^2} \approx q^{n_\text{tot}}e^{-q^2/8}, 
\end{equation}
where the approximation is valid when large squeezing is used. We make this approximation such that the following analysis does not depend on the number of cat amplification rounds or on the squeezing used for cat amplification. To prepare an approximate logical GKP state with peak spacing $2\sqrt{2\pi}$ in $q$ after $N_\text{GKP}$ rounds of the breeding protocol (i.e., consuming $2^{N_\text{GKP}}$ squeezed cats), we consider squeezing the state to have peak spacing $2^{1+N_\text{GKP}/2}\sqrt{2\pi}$. The original states have peaks at $\pm 2\sqrt{n}$, so we squeeze to modify $q$ by a factor $(2\sqrt{n})/(2^{N_\text{GKP}/2}\sqrt{2\pi})$. That is, we consider using the states $\Psi_\textbf{n}( 2^{\frac{1-N}{2}}\sqrt{n/\pi} ~ q)$ as inputs for the GKP breeding protocol. 

\begin{figure}
    \centering
    \includegraphics[width=0.6\linewidth]{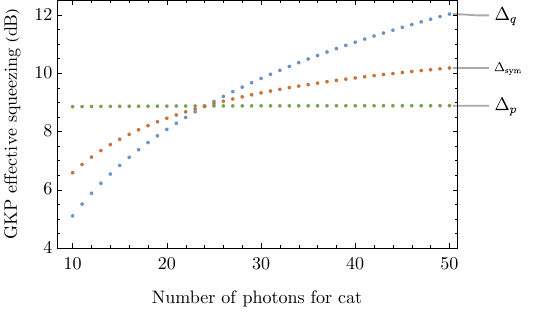}
    \caption{Effective squeezing of approximate GKP states generated from one round of GKP breeding using two cat states. The horizontal axis is the number of photons detected to herald the cat states. The squeezing pre-processing before the GKP breeding protocol, as described in the text, ensures that the approximate GKP states lie on the same grid, regardless of the heralding photon number. }
    \label{fig:effsqz}
\end{figure}

In Fig.~\ref{fig:effsqz}, we plot the effective squeezing metrics for one round of GKP breeding as a function of the number of photons used to prepare input cat states. The symmetric effective squeezing is at least 10 dB when the number of photons is 45, which we use as the number threshold in Fig.~2 of the main text. 

\section{Effective loss simulations}
In the main text, we describe simulations to study the effective amount of loss incurred under different physical losses; here, we provide additional details. For a given output state, we consider the effective loss, $\ell_\text{eff}$, defined by
\begin{equation}
    \ell_\text{eff} = \argmax_{\ell_\text{eff},\alpha,t,r_\text{c}}\mathcal{F}(\text{output},L_{\ell_\text{eff}} [\hat{S}(r_c)\ket{\mathcal{C}(\alpha,t)}), \label{eq:effloss}
\end{equation}
where $\mathcal{F}(\rho_1,\rho_2) = (\Tr\sqrt{\sqrt{\rho_1}\rho_2\sqrt{\rho_1}})^2$ is the fidelity. 

To report a single value of the effective loss for a given set of physical parameters, we determine an expected effective loss, averaged over the distribution of heralded output states. To simplify the simulation, we explicitly select photon-number measurement outcomes, and approximate the average effective loss by pre-selecting the most likely outcomes (see below). We simulate using the \texttt{MrMustard} Python package in a 25-dimensional space, and therefore limit the total number of detected photons to 9 (to keep the output state well-represented in the truncated Hilbert space). The squeezing level for each squeezer is set to 6 dB, and we consider squeezing and PNR detection losses between 0\% and 3\%, in 0.5\% increments. 

The most likely measurement outcomes (excluding detecting zero photons in all rounds) and the corresponding probabilities are determined using Eq.~\eqref{eq:catprob}, and are, in format $([n_1,\dots], p(n_1,\dots))$:
\begin{itemize}[leftmargin=2cm ]
    \item[$N=1$:] ([1], 0.181289), ([2], 0.0717142), ([3], 0.0315207), ([4], 0.0145471), ([5], 0.00690545), ([6], 0.00333869), ([7], 0.00163517), ([8], 0.000808551), ([9], 0.000402769). 

    Corresponds to 99.9\% of all non-zero outcomes in a single round. 
    \item[$N=2$] ([0, 1], 0.132244), ([0, 2], 0.0671862), ([0, 3], 0.0379264), ([1, 1], 0.0337528), ([1, 0], 0.0332182), ([1, 2], 0.02858), ([1, 3], 0.0225867), ([0, 4], 0.0224798), ([1, 4], 0.0172126), ([0, 5], 0.0137049), ([1, 5], 0.0128257), ([1, 6], 0.0094121), ([2, 3], 0.00864723), ([2, 2], 0.00851027), ([0, 6], 0.00851002), ([2, 4], 0.00805419), ([2, 1], 0.00717898), ([2, 5], 0.00709264), ([1, 7], 0.00683113), ([2, 6], 0.00600566), ([0, 7], 0.00535289), ([2, 7], 0.00493996), ([1, 8], 0.00491659), ([2, 0], 0.00423916). 
    
    Corresponds to 95.2\% of outcomes with up to 9 total detected photons.
    
    \item[$N=3$] ([0, 0, 1], 0.0740703), ([0, 0, 2], 0.0405253), ([0, 0, 3], 0.0246358), ([0, 1, 1], 0.020359), ([0, 1, 0], 0.0186056), ([0, 1, 2], 0.0185647), ([0, 1, 3], 0.0157999), ([0, 0, 4], 0.0157252), ([0, 1, 4], 0.0129667), ([0, 1, 5], 0.010405), ([0, 0, 5], 0.0103243), ([0, 1, 6], 0.00822292), ([0, 0, 6], 0.00690386), ([0, 2, 4], 0.00653406), ([0, 2, 3], 0.00651417), ([0, 1, 7], 0.00642703), ([0, 2, 5], 0.00619651), ([0, 2, 2], 0.00595315), ([0, 2, 6], 0.00565039), ([1, 0, 1], 0.00511395), ([0, 2, 7], 0.00500519), ([0, 1, 8], 0.0049815), ([0, 0, 7], 0.00467658), ([1, 0, 0], 0.00467352), ([1, 0, 2], 0.00466324), ([0, 2, 1], 0.00466324). 
    
    Corresponds to 80.6\% of outcomes with up to 9 total detected photons. 
\end{itemize}

Based on the effective loss values calculated using Eq.~\eqref{eq:effloss} for each of the output states, the reported values in Fig.~4(a)--(c) of the main text are the average of the effective losses, weighted by the relative frequency of the measurement pattern (the above probabilities normalised by their sum). In this way, the reported average effective loss approximates the true expected effective loss.

\subsection{Output states from a single round including losses}
Here, we demonstrate that it is possible to exactly calculate the output state for a single round of the protocol including losses. We begin by considering the state heralded by a photon-number measurement on one mode of an arbitrary two-mode Gaussian state. This allows us to model the effects of arbitrary Gaussian noise processes.  

\subsubsection{Conditional states heralded from an arbitrary two-mode Gaussian state}

An arbitrary two-mode Gaussian state has covariance matrix
\begin{equation}
	\Gamma = \begin{pmatrix}
		a &  & e &  \\
		& b & & f \\
		e & & c & \\
		& f & & d
	\end{pmatrix},
\end{equation}
and the corresponding Wigner function is
\begin{equation}
	W_\Gamma(\mathbf{r})= \frac{\exp\left(-\frac{1}{2}(\mathbf{r}-\bar{\mathbf{r}})^\top \Gamma^{-1}(\mathbf{r}-\bar{\mathbf{r}})\right)}{(2\pi)^2\sqrt{\det \Gamma}}, \quad \mathbf{r} = (q_0,p_0,q_1,p_1)^\top. 
\end{equation}
For our purposes, it is sufficient to assume a zero mean vector, $\bar{\mathbf{r}}=0$. Then, the Wigner function is
\begin{align}
	W_\Gamma(\mathbf{r}) &= \frac{1}{4\pi^2\sqrt{\det\Gamma}}\exp\left(-\frac{1}{2}\left(\frac{q_0^2}{a} + \frac{p_0^2}{b}\right)\right)\exp\left(-\frac{1}{2}\left(\frac{a}{\eta_1}\left(\frac{e}{a}q_0-q_1\right)^2 + \frac{b}{\eta_2}\left(\frac{f}{b}p_0-p_1\right)^2\right)\right) \\
	&= \frac{1}{\sqrt{\det\Gamma}}W_0\left(\frac{q_0}{\sqrt{a}},\frac{p_0}{\sqrt{b}}\right)W_0\left(\sqrt{\frac{a}{\eta_1}}\left(\frac{e}{a}q_0-q_1\right), \sqrt{\frac{b}{\eta_2}}\left(\frac{f}{b}p_0-p_1\right)\right),
\end{align}
where
\begin{equation}
	\eta_1=ac-e^2, \quad \eta_2=bd-f^2, \quad \det\Gamma = \eta_1\eta_2.
\end{equation}
For the heralded output state when detecting $\ket{m}$ in mode 1, we calculate $\Tr_1[\hat{\rho}_\Gamma \ketbra{m}]$. The associated probability is $\Tr_{0,1}[\hat{\rho}_\Gamma\ketbra{m}]$. In terms of the Wigner functions, the partial trace is
\begin{align}
	\Tr_1[\hat{\rho}_\Gamma\ketbra{m}] &= 4\pi\iint \mathrm{d}q_1\mathrm{d}p_1~W_\Gamma(q_0,p_0,q_1,p_1)W_m(q_1,p_1) \\
	&= \frac{4\pi}{\sqrt{\det\Gamma}}W_0\left(\frac{q_0}{\sqrt{a}},\frac{p_0}{\sqrt{b}}\right)\iint \mathrm{d}q_1\mathrm{d}p_1~W_0\left(\sqrt{\frac{a}{\eta_1}}\left(\frac{e}{a}q_0-q_1\right), \sqrt{\frac{b}{\eta_2}}\left(\frac{f}{b}p_0-p_1\right)\right)W_m(q_1,p_1).
\end{align}
We therefore require integrals of the form
\begin{align}
	I_1(q_0,p_0,m,\alpha,\beta) &= \iint \mathrm{d}q_1\mathrm{d}p_1~W_0(\alpha(q_0-q_1)),\beta(p_0-p_1))W_m(q_1,p_1) \\
	&= \frac{(-1)^m}{(2\pi)^2}\iint \mathrm{d}q_1\mathrm{d}p_1~e^{-(\alpha^2(q_0-q_1)^2+\beta^2(p_0-p_1)^2)/2}e^{-(q_1^2+p_1^2)/2}L_m(q_1^2+p_1^2).
\end{align}

To evaluate this integral, we can use the generating function of the Laguerre polynomials:
\begin{equation}
	\sum_{n=0}^\infty t^n L_n(q) = \frac{1}{1-t}e^{-tz/(1-t)}.
\end{equation}
The desired integrals will then be the power series coefficients of the auxiliary integral $I_2 = \sum_{n=0}^\infty t^nI_1(n)$,
\begin{align}
	I_2(q_0,p_0,t,\alpha,\beta) &= \frac{(-1)^m}{(2\pi)^2}\frac{1}{1-t}\iint \mathrm{d}q_1\mathrm{d}p_1~e^{-(\alpha^2(q_0-q_1)^2+\beta^2(p_0-p_1)^2)/2}e^{-(q_1^2+p_1^2)/2}e^{-\frac{t}{1-t}(q_1^2+p_1^2)} \\
	&= \frac{(-1)^m}{2\pi}\frac{e^{-\left(\frac{(1+t)\alpha^2}{1+\alpha^2+t(1-\alpha^2)}\frac{q_0^2}{2}+\frac{(1+t)\beta^2}{1+\beta^2+t(1-\beta^2)}\frac{p_0^2}{2}\right)}}{\sqrt{1+\alpha^2+t(1-\alpha^2)}\sqrt{1+\beta^2+t(1-\beta^2)}}.
\end{align}
Then we can calculate
\begin{align}
	I_1(q_0,p_0,0,\alpha,\beta) &= \frac{e^{-\left(\frac{\alpha^2}{1+\alpha^2}q_0^2+\frac{\beta^2}{1+\beta^2}p_0^2\right)/2}}{2\pi \sqrt{(1+\alpha^2)(1+\beta^2)}}, \\
	I_1(q_0,p_0,1,\alpha,\beta) &= I_1(q_0,p_0,0,\alpha,\beta)\left(\frac{\alpha^4q_0^2-\alpha^4-\alpha^2}{(1+\alpha^2)^2}+\frac{\beta^4p_0^2+\beta^2+1}{(1+\beta^2)^2}\right),
\end{align}
and so on. Now we can evaluate the conditional (unnormalised) Wigner functions as
\begin{equation}
	W_{\Gamma,m}(q_0,p_0) = \frac{4\pi}{\sqrt{\det\Gamma}}W_0\left(\frac{q_0}{\sqrt{a}},\frac{p_0}{\sqrt{b}}\right)I_1\left(\frac{e}{a}q_0,\frac{f}{b}p_0, m, \sqrt{\frac{a}{\eta_1}},\sqrt{\frac{b}{\eta_2}}\right). \label{eq:condwig}
\end{equation}
To get the detection probabilities, we integrate over $q_0,p_0$. The result is the series expansion coefficients of the function
\begin{align}
	I_3(\Gamma,m,t) &= \frac{4\pi}{\sqrt{\det\Gamma}}\iint \mathrm{d}q_0\mathrm{d}p_0~W_0\left(\frac{q_0}{\sqrt{a}},\frac{p_0}{\sqrt{b}}\right)I_2\left(\frac{e}{a}q_0,\frac{f}{b}p_0, t, \sqrt{\frac{a}{\eta_1}},\sqrt{\frac{b}{\eta_2}}\right) \\
	&= \frac{2(-1)^m}{\sqrt{(1+c+t(c-1))(1+d+t(d-1))}}.
\end{align}
Then, the probabilities are defined recursively by
\begin{align}
	p(0) &= \frac{2}{\sqrt{(1+c)(1+d)}},\\
	p(1) &= \frac{2(cd-1)}{((1+c)(1+d))^{3/2}}, \\
	p(m+2) &= \frac{1}{(2+m)(1+c)(1+d)}\left((3+2m)(cd-1)p(m+1)-(1+m)(c-1)(d-1)p(m)\right), 
	\\ p(2) &= \frac{2-6 c+d^2+c^2(1+2d^2)}{((1+c)(1+d))^{5/2}} \\
	p(3) &= \frac{(cd-1)(2-10cd+3d^2+c^2(3+2d^2))}{((1+c)(1+d))^{7/2}}
\end{align}

\subsubsection{Fidelity for single-round cat states including losses}

\begin{figure*}
	\includegraphics{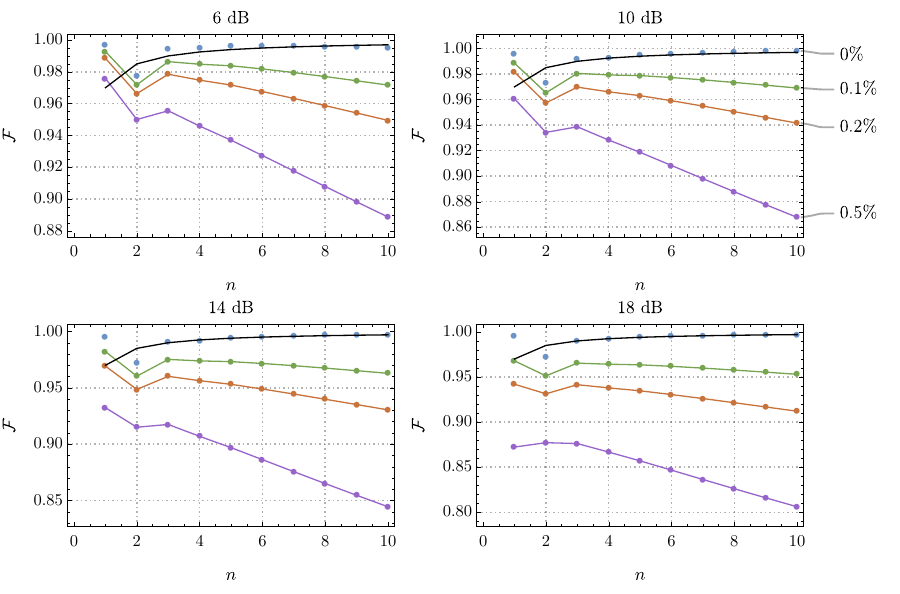}
	\caption{Fidelity of output states with respect to a cat state $\ket{\mathcal{C}(\sqrt{n},n)}$, for different squeezing levels and including loss. Pure loss with strength $\ell$ is applied after each squeezer and before the PNR detector, and values $\ell=0$, $0.1\%$, $0.2\%$, $0.5\%$ are shown. The black line is $\mathcal{F}=1-0.03/n$, which approximates the fidelity for $n>1$ and large squeezing. Fidelities calculated numerically based on the Wigner function of the output and target state.}
	\label{fig:fidlossy}
\end{figure*}

We can compare the conditional output Wigner functions from Eq.~\eqref{eq:condwig} to ideal squeezed cat states. For cat states of the form 
\begin{equation}
	\hat{S}(r)\ket{\text{cat}_{\alpha,k}} = \mathcal{N}\hat{S}(r)\left(\ket{\alpha} + (-1)^k\ket{-\alpha}\right)
\end{equation}
with real $\alpha$ and $r$, the normalised $q$-quadrature wavefunction is
\begin{equation}
	\bra{q}\hat{S}(r)\ket{\text{cat}_{\alpha,k}} = \frac{e^{r/2}}{(2\pi)^{1/4}\sqrt{2+2(-1)^ke^{-2\alpha^2}}}\left(e^{-(e^rq-2\alpha)^2/4}+(-1)^ke^{-(e^{r}q+2\alpha)^2/4}\right).
\end{equation}
The corresponding Wigner function is
\begin{equation}
	W^\text{cat}(q,p,\alpha,r,k) = \frac{e^{-\frac{1}{2} \left(p^2 e^{-2 r}+q^2 e^{2 r}+4 \alpha  q e^r\right)} \left(2 (-1)^k e^{2 \alpha  \left(\alpha +q e^r\right)} \cos \left(2 \alpha  p e^{-r}\right)+e^{4 \alpha  q e^r}+1\right)}{4 \pi  \left(e^{2 \alpha ^2}+(-1)^k\right)}.
\end{equation}
As the target state is pure, we can calculate the fidelity as $F = 4\pi \iint \mathrm{d}q\mathrm{d}p~W^\text{cat}(q,p,\alpha,r,k)W_{\Gamma,m}(q,p)$. As an example, in Fig.~\ref{fig:fidlossy}, we present the fidelity of the output states including loss with respect to a cat state $\ket{\mathcal{C}(\sqrt{n},n)}$. Note that the fidelity with respect to a different cat state may be higher. 

\section{Required squeezing for equivalent gaussian boson sampling circuit}
In the main text, we describe how our repeated circuit (as in Fig.~1(a)) can be rearranged into an equivalent circuit that requires only offline squeezers and a linear interferometer (hereafter referred to as ``GBS circuit''). The important question is: how much offline squeezing is required for the GBS circuit equivalent to $N$ rounds of the repeated scheme using a certain amount of inline squeezing (per squeezer)? Intuitively, the offline squeezing increases with $N$ because we are repeatedly squeezing the signal mode. 

To determine the required offline squeezing, we calculate the singular values of the symplectic matrix that transforms (both $q$ and $p$ quadratures) from the $(N+1)$-mode vacuum into the state of the multiple-round scheme before photon counting (deferring all photon-number measurements to the end). These singular values determine the amount of squeezing. Here, we restrict attention to the \texttt{cat} scheme, but the results can be easily generalised to the \texttt{squeezed-cat} scheme.

\begin{figure}
    \centering
    \includegraphics[width=\linewidth]{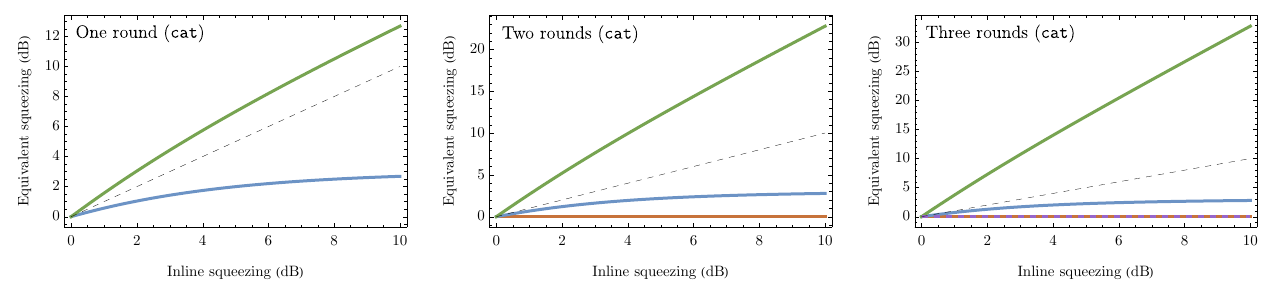}
    \caption{Required offline squeezing for equivalent GBS circuit for \texttt{cat} scheme. The horizontal axis is the inline squeezing in dB, and the vertical axis is the equivalent offline squeezing in dB. The different lines represent the $N+1$ different inputs required for the GBS circuit that is equivalent to $N$ rounds; for three rounds, there are two vacuum inputs. }
    \label{fig:GBSsqueeze}
\end{figure}

As before, we determine the transformation matrices by iterative application of a matrix that couples the signal to each ancilla. For a single round, we have 
\begin{align}
\begin{split}
	O^{(1)} &= \begin{pmatrix}
		e^{r} & 0 & 0 & 0 \\
		0 & e^{-r} & 0 & 0 \\
		0 & 0 & 1 & 0 \\
		0 & 0 & 0 & 1
	\end{pmatrix}\begin{pmatrix}
		-\sqrt{R'} &  0 & \sqrt{1-R'} & 0 \\
		0 & -\sqrt{R'} & 0 & \sqrt{1-R'} \\
		\sqrt{1-R'} & 0 & \sqrt{R'} & 0 \\
		0 & \sqrt{1-R'} & 0 & \sqrt{R'}
	\end{pmatrix}\\ &\qquad \quad \times \begin{pmatrix}
		e^{-r} & 0 & 0 & 0 \\
		0 & e^{r} & 0 & 0 \\
		0 & 0 & e^r & 0 \\
		0 & 0 & 0 & e^{-r}
	\end{pmatrix}\begin{pmatrix}
		\sqrt{R} &  0 & \sqrt{1-R} & 0 \\
		0 & \sqrt{R} & 0 & \sqrt{1-R} \\
		-\sqrt{1-R} & 0 & \sqrt{R} & 0 \\
		0 & -\sqrt{1-R} & 0 & \sqrt{R}
	\end{pmatrix}
	\end{split} \\
	& = \begin{pmatrix}
		-e^{r} & 0 & 0 & 0 \\
		0 & -e^{-r} & 0 & -1+e^{-2r} \\
		e^{-r}-e^r & 0 & 1 & 0 \\
		0 & 0 & 0 & 1
	\end{pmatrix}.
\end{align}
For additional rounds, because each round only couples the signal to a single ancilla, the transformation matrices have the same form, i.e., for two rounds the overall transformation is given by the product 
\begin{align}
	O^{(2)}&= O_2O_1 = \begin{pmatrix}
		-e^{r} & 0 & 0 & 0 & 0 & 0\\
		0 & -e^{-r} & 0 &0 & 0& -1+e^{-2r}  \\
		0 & 0 & 1 & 0 & 0 & 0 \\
		0 & 0 & 0 & 1 & 0 & 0 \\
		e^{-r}-e^r & 0 & 0 & 0 & 1 & 0 \\
		0 & 0 & 0 & 0 & 0 & 1
	\end{pmatrix}\begin{pmatrix}
		-e^{r} & 0 & 0 & 0 & 0 & 0  \\
		0 & -e^{-r} & 0 & -1+e^{-2r} & 0 & 0  \\
		e^{-r}-e^r & 0 & 1 & 0  & 0 & 0 \\
		0 & 0 & 0 & 1 & 0 & 0 \\
		0 & 0 & 0 & 0 & 1 & 0 \\
		0 & 0 & 0 & 0 & 0 & 1
	\end{pmatrix} \\
	&= \begin{pmatrix}
		e^{2r} & 0 & 0 & 0& 0 & 0 \\
		0 & e^{-2r} & 0 & e^{-3r}(e^{2r}-1) & 0 & e^{-2r}-1 \\
		e^{-r}-e^r & 0 & 1 & 0 & 0 & 0  \\
		0 & 0 & 0 & 1 & 0 & 0 \\
		-1+e^{2r} & 0 & 0 & 0 & 1 & 0 \\
		0 & 0 & 0 & 0 & 0 & 1 
	\end{pmatrix}
\end{align}
This can be repeated for the desired number of rounds. Finally, we determine the singular values of $O^{(N)}$ to determine the equivalent amount of squeezing. A sample is presented in Fig.~\ref{fig:GBSsqueeze}, where we represent the squeezing in dB by calculating $20\log_{10}(s_i)$, where $s_i$ is a singular value of $O^{(N)}$ (equivalently, we could calculate $10\log_{10}(\lambda_i)$, where $\lambda_i$ is an eigenvalue of $O^{(N)}(O^{(N)})^\top$). As the singular values come in pairs (corresponding to squeezing and anti-squeezing), we present only the positive values. We find that not all inputs are required to be squeezed, but as the number of rounds is increased, the maximum squeezing increases quickly.  

\bibliography{bib.bib}